
\magnification=\magstep1
\hfuzz=4.pt
\baselineskip 20 pt
\hsize=15truecm
\vsize=23truecm
\voffset=0pt
\hoffset=1truecm
\def\big{\bigskip \noindent}
\def\med{\medskip \noindent}
\def\c{\line}
\def\i{\item}

\parindent=0mm
\centerline{\bf Hidden Errors in Simulations and the Quality of Pseudorandom
                Numbers}

\big
\big
In a recent letter Ferrenberg {\it et al.} (FLW) [1] present intriguing
results arising from combinations of some random number generators
and Monte Carlo acceleration algorithms. In particular, they observe
systematic errors when the Wolff algorithm [2] is used at $T_c$ in the
$2-d$ Ising model.
As an explanation of this, they propose that subtle correlations
arise in the random number sequences, in the sense that the
higher order bits of the random numbers are correlated.

\med
We have recently carried out extensive statistical,
bit level, and visual tests for several commonly used
pseudorandom number generators in physics applications [3].
Two of these were used in Ref. 1, namely
R250 [4] and CONG [5].
Using the Wolff algorithm FLW discovered problems
with R250, but not with CONG. Thus, our test results bear
direct relevance to the existence of possible problems, and
differences between these two algorithms.

\med
To directly probe the correlations between each bit of consecutive pseudorandom
numbers, we have performed two quantitative tests. The first one
is an extended version of the $m$-tuple test [6].
The second test was based on the range of binary
random matrices [7].
In Table 1 we present a summary of the results
for each of the 31 significant bits from random
numbers generated by R250 and CONG.
Our results indicate no bit level correlations in CONG.
The same is true for R250, {\it provided} it is properly initialized.
This is demonstrated in Table 2, where R250 was initialized by a
lagged Fibonacci generator RAN3 [8], which contains several
correlated bits, including five of the most significant ones.
As seen in the results, correlations from RAN3 transform into R250.

\med
Our results thus indicate that the problems observed
in Ref. 1 have no simple explanation in terms of bit level correlations.
We note that both R250 and CONG were also subjected to an array of
statistical tests [3,5], in which neither of them
showed any particular weaknesses. However, our $2-d$ visual tests
did reveal a periodic spatial fine structure in CONG,
a result not unexpected with this type of algorithm [3].
Yet, no problems with CONG were reported by FLW.
Since their results presently have no clear explanation,
we wholeheartedly agree with them
on the importance of careful physical tests [3]
before a particular "good quality" generator is chosen for simulations.

\big
\big
\c{I. Vattulainen$^1$, K. Kankaala$^{1,2}$, J. Saarinen$^1$, and
T. Ala-Nissila$^3$ \hfil}

\baselineskip=12pt

\big
\c{$^1$Department of Electrical Engineering \hfil}
\c{Tampere University of Technology \hfil}
\c{P.O. Box 692 \hfil}
\c{SF-33101 Tampere \hfil}
\c{Finland \hfil}

\med
\c{$^2$CSC Scientific Computing \hfil}
\c{P.O. Box 405 \hfil}
\c{SF-02101 Espoo \hfil}
\c{Finland \hfil}

\med
\c{$^3$Research Institute for Theoretical Physics \hfil}
\c{P.O. Box 9 (Siltavuorenpenger 20 C) \hfil}
\c{SF-00014 University of Helsinki \hfil}
\c{Finland \hfil}

\big

\vfill \eject

\baselineskip=16pt
\centerline{\bf References:}

\big
\i{[1]} A. M. Ferrenberg, D. P. Landau, and Y. J. Wong, Phys. Rev. Lett.
       {\bf 69}, 3382 (1992).

\i{[2]} U. Wolff, Phys. Rev. Lett. {\bf 62}, 361 (1989).

\i{[3]} I. Vattulainen, K. Kankaala, J. Saarinen, and T. Ala-Nissila,
        CSC Research Report R05/92 (Centre for Scientific Computing,
        Espoo, Finland, 1992); to be published.

\i{[4]} S. Kirkpatrick and E. Stoll, J. Comput. Phys. {\bf 40}, 517 (1981).

\i{[5]} D. E. Knuth, {\it The Art of Computer Programming, vol. 2:
Seminumerical
        Algorithms}, 2nd. ed. (Addison - Wesley, Reading, 1981).

\i{[6]} S. N. Altman, J. Sci. Stat. Comput. {\bf 9}, 941 (1988).

\i{[7]} G. A. Marsaglia, in {\it Computational Science and Statistics:
        The Interface}, ed. L. Billiard (Elsevier, New York, 1985).

\i{[8]} W. H. Press, B. P. Flannery, S. A. Tenkolsky, and W. T. Vetterling,
        {\it Numerical Recipes: The Art of Scientific Computing
	(FORTRAN Version) } (Cambridge University Press, 1989).

\vfill \eject
\hoffset=-1truecm \hsize=12truecm \centerline{\bf Table Captions:}
\big
\big
\rm
\vbox{\tabskip=0pt \offinterlineskip
\halign{
\strut \vrule \quad #\quad & \vrule \hfil \quad #\quad \hfil &
\vrule \hfil \quad #\quad \hfil \vrule \cr
\noalign{\hrule}
Random & Failing bits & Failing bits \cr
number & in the & in the \cr
generator & {\it m}-tuple test & random matrix test \cr
\noalign{\hrule}
R250 & none & none \cr
CONG & none & none \cr
\noalign{\hrule}
}
}
\hoffset=3truecm
\hsize=12truecm
\big
\i{Table 1.} The \hfil results \hfil of \hfil an \hfil extended \hfil
{\it m}-tuple \hfil test \hfil with \hfil parameters \break
$m~=~t~=~3$, and a test using $2 \times 2$ random matrices. In these tests
R250 was initialized by CONG.
\big
\big
\big
\big
\vbox{\tabskip=0pt \offinterlineskip
\halign{
\strut \vrule \quad #\quad & \vrule \hfil \quad #\quad \hfil &
\vrule \hfil \quad #\quad \hfil \vrule \cr
\noalign{\hrule}
Random & Failing bits & Failing bits \cr
number & in the & in the \cr
generator & {\it m}-tuple test & random matrix test \cr
\noalign{\hrule}
R250 & 1 - 2, 27 - 31 & 1, 27 - 31 \cr
RAN3 & 1 - 5, 25 - 30 & 1 - 5, 26 - 30 \cr
\noalign{\hrule}
}
}
\big
\i{Table 2.} The results of bit level tests for R250 initialized by RAN3.

\bye